# Tracking the Evolution of Near-Field Photonic Qubits into High-Dimensional Qudits via State Tomography

Amit Kam[1,2], Shai Tsesses[3,4], Lior Fridman[2,3], Yigal Ilin[2,3], Amir Sivan[2,3], Guy Sayer[2,3], Kobi Cohen[3,5], Amit Shaham[3], Liat Nemirovsky-Levy[1,2,5], Larisa Popilevsky[6], Meir Orenstein[2,3,6], Mordechai Segev[1,2,3,5] and Guy Bartal[2,3,6*]

[1]*Physics department, Technion – Israel Institute of Technology, Haifa 32000, Israel*
[2]*Helen Diller Quantum Center, Technion – Israel Institute of Technology, Haifa 32000, Israel*
[3]*Andrew and Erna Viterbi department of Electrical & Computer Engineering, Technion – Israel Institute of Technology, Haifa 32000, Israel*
[4]*Department of Physics, MIT-Harvard Center for Ultracold Atoms and Research Laboratory of Electronics, Massachusetts Institute of Technology, Cambridge, MA 02139, USA*
[5]*Solid state institute, Technion – Israel Institute of Technology, Haifa 32000, Israel*
[6]*Russell Berrie Nanotechnology Institute, Technion, Haifa 32000, Israel*
Author e-mail address: *guy@ee.technion.ac.il

**Abstract**

**Quantum nanophotonics offers essential tools and technologies for controlling quantum states, while maintaining a miniature form factor and high scalability. For example, nanophotonic platforms can transfer information from the traditional degrees of freedom (DoFs), such as spin angular momentum (SAM) and orbital angular momentum (OAM), to the DoFs of the nanophotonic platform - and back, opening new directions for quantum information processing. Recent experiments have utilized the total angular momentum (TAM) of a photon as a unique means to produce entangled qubits in nanophotonic platforms. Yet, the process of transferring the information between the free-space DoFs and the TAM was never investigated, and its implications are still unknown. Here, we reveal the evolution of quantum information in heralded single photons as they couple into and out of the near-field of a nanophotonic system. Through quantum state tomography, we discover that the TAM qubit in the near-field becomes a free-space qudit entangled in the photonic SAM and OAM. The extracted density matrix and Wigner function in free-space indicate state preparation fidelity above 97%. The concepts described here bring new concepts and methodologies in developing high-dimensional quantum circuitry on a chip.**

**Main**

Quantum optics at the nanoscale explores the enhanced interaction between photons and emitters at the sub-wavelength level[1,2], as well as the possibilities to engineer the optical environment of photons via nano-structuring to produce new quantum states of light[3,4] for quantum communications[5–7], sensing[8,9] and computation[10–12] within compact dimensions and in a scalable chip-based platform[13,14]. In this vain, photons are versatile carriers of quantum information, capable of encoding qubits across a variety of degrees-of-freedom (DoFs), including propagation direction[15], time-[16] or frequency-bin[17], spatial mode shape[18–21] and polarization[22–24]. The latter two DoFs can both be related to angular momentum: an angular phase gradient leads to orbital angular momentum (OAM), while circular polarization manifests the spin angular momentum (SAM) of a photon[25–27].

These DoFs are separable when the shape of the photonic wavepacket varies slowly on the wavelength scale, as in the paraxial approximation. The ability to separate the SAM and the OAM of a photon enables entanglement between these DoFs, allowing for more quantum information to be encoded on individual photons[28,29]. However, when photons are tightly confined on the scale of their wavelength, the SAM and OAM become inseparable[30,31], leaving the total angular momentum (TAM) as the characteristic quantum number[32–35]. Entanglement in the TAM DoF has recently been demonstrated in nanophotonic systems[36], and that initial exploration indicated that the coupling of the photons out to free-space can potentially generate *qudits* - quantum units existing in a *d*-dimensional Hilbert space, instead of the two-dimensional space of qubits.

Here, we employ the transfer between DoFs in tightly confined photons in nanophotonic platform to generate mode-polarization *qudits*. These qudits are generated by coupling incident heralded single photons into surface plasmon polariton (SPP) excitations, with their near-field mode forming TAM qubits. These qubits are then scattered out into free space where they become single photonic qudits, entangled in their SAM and OAM. We perform Quantum state tomography (QST) on the free-space qudits to unravel the quantum information held by near-field qubits, characterizing the resulting qudit basis through projections onto all SAM-OAM combinations. By reconstructing the photonic density matrix and Wigner function, we identify that the state preparation fidelity of the near-field TAM qubits is above 97%. This study highlights the unique evolution of the quantum information held by a single photon as it propagates through a nanophotonic platform, paving the way for novel on-chip implementation of quantum circuity and an on-chip source for qudits with judiciously engineered properties.

**TAM Integration in Nanophotonics**

We employ a plasmonic platform (Fig. 1a) consists of a gold-air interface patterned with a circular grating to couple light into SPP modes $|J_n\rangle$ carrying angular momentum [30,37–40]. In such systems, the TAM of the photon eigenvalues $j = n\hbar$ cannot be simply separated into independent measurable components of proper SAM and OAM[41]. Instead, any near-field state is a vector electric field mode, which can be expressed in terms of cartesian X, Y, and Z components, or as a combination of in-plane rotating field components $(LH = \frac{X+iY}{2}, RH = \frac{X-iY}{2})$ along with its out-of-plane z-component. Each field component has its own spatial distribution.

In particular, the in-plane rotating components of a nanophotonic mode generated by an incident circularly polarized photon can be described by Bessel-type modes, while a linearly polarized input transforms into a superposition of Bessel modes, more naturally represented by Hermite-Bessel-type modes, which lack the rotational symmetry [42–45].

An additional concentric annular slit is carved half-way into the gold, coupling the SPPs back into free-space photons. As we recently shown[36], the circular symmetry of this system preserves the angular momentum properties in the out-coupling process, facilitating the study of angular momentum transfer in nanoscale photonic systems.

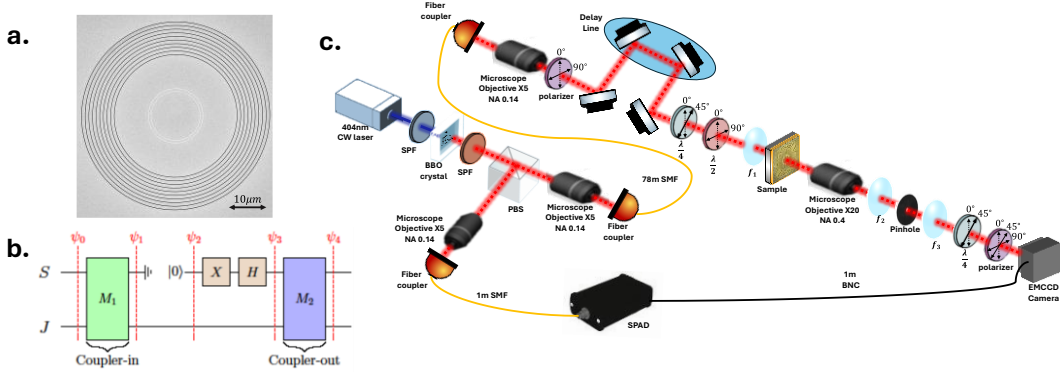

**Fig. 1. Experimental apparatus for heralded detection of nanophotonic modes**: **a.** SEM micrograph of the nanophotonic platform - a nanopatterned gold layer evaporated on a glass substrate. The circular input coupler, which couples photons of a given polarization to plasmons with a well-defined TAM, is milled through the entire gold layer. The annular out-coupler ring is milled only through half of it, scattering the SPPs towards the camera. **b.** Schematic representation of the nanophotonic platform as a quantum circuit utilizing two DoFs of the launched photon. (See supplementary for the derivation as a quantum channel). **c.** Experimental setup: a 404nm CW laser is directed into a BBO type- II crystal to produce photon pairs via spontaneous parametric down conversion (SPDC). The photon pairs are filtered to allow only collinear, same-frequency photons at 808nm, distinguishable only by their orthogonal linear polarizations. The photons are separated into two different paths using a polarizing beam splitter (PBS). The first (heralding) photon is coupled into a single-mode fiber and sent to a single-photon detector, whose signal triggers the EMCCD camera and ensures it only detects coincident single photon events. The second photon is coupled into a different single-mode fiber, then enters a delay line to synchronize the timing between the heralding signal and the photon arrival on the EMCCD camera. After passing through waveplates to control its polarization, the photon is incident upon the nanophotonic platform (denoted as 'sample'). Once scattered, the outgoing photon passes through waveplates to project the quantum state onto different polarization components.

In general, SPPs generated in such circularly-symmetric nanophotonic platform are either eigenstates or coherent superpositions of the TAM operator, with a TAM value equal to that of the incident free space photon whose TAM is the sum of its SAM and OAM values[36]. This eigenstate corresponds to the $n^{th}$ order transverse magnetic (TM) Bessel mode ($|J_n\rangle$ carrying $j = n\hbar$ angular momentum). It is important to note that the mode in the near-field is insensitive to the specific angular momentum decomposition of incident photons: two different free-space photonic states (e.g., $|\sigma_+\rangle \otimes |l = 0\rangle$ and $|\sigma_-\rangle \otimes |l = 2\rangle$) can excite the same SPP mode with TAM of $j = 1$. We generate such states by launching onto the nanophotonic platform paraxial photons carrying only SAM. $\sigma\hbar$ corresponds to the SAM associated with left ($\sigma = -1$) and right ($\sigma = +1$) handed circularly polarized modes $|\sigma_+\rangle$, and $|\sigma_-\rangle$ and $l = n\hbar$ where $n$ is an integer are the OAM eigenvalues manifested by an azimuthal phase gradient $\exp(in\phi)$ of the field. The incident photons are either circular polarization states or their superpositions, hence they transform into SPP modes with a well-defined TAM (or their superposition) per photon, which is equal to the SAM of the incident photon. Specifically, a $|\sigma_-\rangle$ ($|\sigma_+\rangle$) right (left) handed circular polarization generates a nanophotonic state of $|J_{-1}\rangle$ ($|J_{+1}\rangle$). Namely, the nanophotonic mode is shaped as a Bessel function of the first kind of order $-1$ ($+1$), carrying a $-\hbar$ ($+\hbar$) quantum of angular momentum. Likewise, the linear polarization states $|H\rangle, |V\rangle$, which are superpositions of circular polarizations ($|H\rangle = \frac{|\sigma_+\rangle + |\sigma_-\rangle}{\sqrt{2}}, |V\rangle = \frac{|\sigma_+\rangle - |\sigma_-\rangle}{\sqrt{2}i}$), generate a superposition of SPP modes $|J_\pm\rangle = \frac{1}{\sqrt{2}}(|J_1\rangle \pm |J_{-1}\rangle)$. Equation (1) summarizes the excitation of these four possible nanophotonic modes as a function of polarization of the incident photon:

$$
\begin{aligned}
|\sigma_+\rangle \otimes |l=0\rangle &\quad\mapsto |J_1\rangle \\
|\sigma_-\rangle \otimes |l=0\rangle &\quad\mapsto |J_{-1}\rangle \\
|H\rangle \otimes |l=0\rangle = \frac{|\sigma_+\rangle + |\sigma_-\rangle}{\sqrt{2}} \otimes |l=0\rangle &\quad\mapsto |J_+\rangle \\
|V\rangle \otimes |l=0\rangle = \frac{|\sigma_+\rangle - |\sigma_-\rangle}{\sqrt{2}i} \otimes |l=0\rangle &\quad\mapsto |J_-\rangle
\end{aligned}
\qquad (1)
$$

where $\mapsto$ indicates the in-coupling operation. The ideal coupling of the input field into the nanophotonic mode facilitates the generation of a qubit encoded in the TAM basis. The generated plasmons are scattered out by the annular coupler. The transversal nature of propagating electromagnetic waves results in scattering of the in-plane components of the SPP field, projecting the eigenstate of the TAM operator in the near-field onto the eigenstate of the TAM in free space (we ignore here the reflected projections). However, in the far-field of the annular coupler, the angular momentum of the scattered field can be decomposed to its SAM and OAM constituents, since the paraxial approximation holds. This gives rise to a unique transformation of the TAM of the near-field photon, embedded in its spatial distribution $|J_n\rangle$, to SAM-OAM entanglement of the free-space photonic state:

$$
|J_n\rangle \mapsto \frac{1}{\sqrt{2}}(|\sigma_-\rangle|l=n+1\rangle - |\sigma_+\rangle|l=n-1\rangle) \qquad (2)
$$

where $\mapsto$ represents here out-coupling operation and $n = (-1,1)$ in our system. Note that for $n = 0$, the nanophotonic mode is purely radial and lacks any azimuthal dependence and thus the state transforms as: $|J_0\rangle \mapsto \frac{1}{\sqrt{2}}(|\sigma_+\rangle|l=-1\rangle - |\sigma_-\rangle|l=1\rangle)$.

This entanglement is a unique result of the fact that the angular momentum of the vector SPP mode is solely characterized by the observable TAM, and that only two combinations of SAM and OAM can satisfy Maxwell's equations when light is coupled to the far-field. Since there is no preferred combination due to the circular symmetry, the scattered photon is in a superposition of all combinations for the given TAM. considering Eqs. (1) and (2) together, out-coupling from the nanophotonic platform yields the resulting states:

$$
\begin{aligned}
|J_1\rangle &\mapsto \frac{1}{\sqrt{2}}(|\sigma_-\rangle|l=2\rangle - |\sigma_+\rangle|l=0\rangle) \\
|J_{-1}\rangle &\mapsto \frac{1}{\sqrt{2}}(|\sigma_-\rangle|l=0\rangle - |\sigma_+\rangle|l=-2\rangle) \\
|J_+\rangle = \frac{1}{\sqrt{2}}(|J_1\rangle + |J_{-1}\rangle) &\mapsto \frac{1}{\sqrt{2}}(|H\rangle|HB_{20}\rangle - |V\rangle|HB_{11}\rangle) \\
|J_-\rangle = \frac{1}{\sqrt{2}}(|J_1\rangle - |J_{-1}\rangle) &\mapsto \frac{1}{\sqrt{2}}(|H\rangle|HB_{11}\rangle - |V\rangle|HB_{02}\rangle)
\end{aligned}
\qquad (3)
$$

where $|HB_{11}\rangle = i\frac{|l=-2\rangle - |l=2\rangle}{\sqrt{2}}$, $|HB_{20}\rangle = \frac{|l=2\rangle + 2|l=0\rangle + |l=-2\rangle}{2}$ and $|HB_{02}\rangle = \frac{-|l=2\rangle + 2|l=0\rangle - |l=-2\rangle}{2}$ are Hermite-Bessel modes. Evidently, starting with incident circularly-polarized photons results in scattered single far-field photons entangled in their SAM and OAM. Similarly, launching linearly polarized photons still results in entanglement of the scattered single photons, but in different polarization and mode bases which has a larger Hilbert space when projected onto SAM and OAM. Therefore, they can be thought of as qudits, with the logical base definition: $|\sigma_+\rangle|l=0\rangle = |1\rangle, |\sigma_-\rangle|l=0\rangle = |2\rangle$, $|\sigma_+\rangle|l=-2\rangle = |3\rangle, |\sigma_-\rangle|l=2\rangle = |4\rangle$. Correspondingly, a qudit base can be defined with linear polarizations and Hermite-Bessel modes for photons entangled in their SAM and OAM: $|H\rangle|HB_{20}\rangle = |\tilde{1}\rangle, |V\rangle|HB_{11}\rangle = |\tilde{2}\rangle, |H\rangle|HB_{11}\rangle = |\tilde{3}\rangle, |V\rangle|HB_{02}\rangle = |\tilde{4}\rangle$.

It is convenient to model the nanophotonic system as a non-unital quantum channel represented by a non-unitary quantum circuit, as shown in Fig. 1b which has been studied in the context of learning local quantum channels[46]. The circuit acts on a composite system of two qubits, utilizing the photon's TAM and SAM degrees of freedom. The output of the quantum circuit in Fig. 1c is an entangled state between the photon's polarization and spatial distribution, with the input and output couplers, $\widehat{M}_1$ and $\widehat{M}_2$, acting as two-qubit unitary gates, while $\hat{X}$ and $\hat{H}$ represent the Pauli $\hat{X}$ and Hadamard operators, respectively. The reduction in degrees of freedom in the near-field is modeled as a dissipative operation, where the action on a single-qubit state $|\psi\rangle$ is represented as $|\psi\rangle \to |0\rangle$. Consequently, the quantum system is both dissipative and entangling. Such dissipative and entangling circuits have recently been shown to enable methods for preparing a wide range of ground states, facilitating the representation of nontrivial thermal states with significantly enhanced noise resilience[47], and able to eliminate the barren plateau effect in the training of quantum neural networks[48].

**Quantum state tomography**

To fully utilize the extended high-dimensional Hilbert space of the emitted qudits, the evolution of the quantum states in the system needs to be thoroughly analyzed and understood. Typically, such a feat is accomplished through Quantum State Tomography (QST)[49,50], which has been used to characterize many different quantum systems[51,52], in particular the quantum states of light[53,54]. QST determines the quantum-mechanical state of a system by measuring various non-commuting observables on an ensemble of identically-prepared system copies, enabling the extraction of the density matrix in a predetermined fashion. We perform quantum state tomography of the photonic qudits generated by passing through our entire apparatus (fig. 1c). We use heralded detection, separating two photons generated via spontaneous parametric down-conversion in a nonlinear crystal and using one of them to trigger an Electron-Multiplying CCD (EMCCD) camera. Hence, the camera only captures single photons scattered from the nanophotonic platform. The imaging system is set in the Fourier plane of the platform, allowing it to image the shape of the modes. On average, only one pixel is turned on by a single photon, and the camera records its position in the image plane of the structure output. We launch single photons into the platform at 4 different polarizations ($|H\rangle, |V\rangle, |\sigma_-\rangle, |\sigma_+\rangle$) using a quarter- (QWP) and a half-wave plate (HWP) and use an additional set of waveplates and a polarizer to project the quantum state of the scattered photon onto these four polarization components, resolving their distribution via the camera.

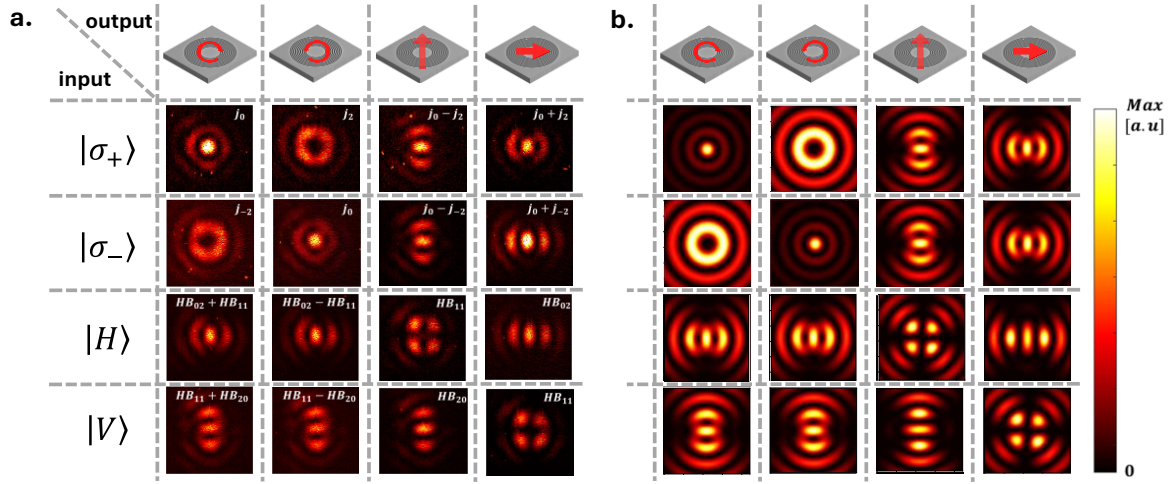

**Fig. 2. Quantum state tomography of photonic free-space qudits created from near-field qubits.** The panels show the intensity of the projection of a heralded single photon on the four different polarizations, as recorded by an EMCCD camera. This action is performed for four different polarizations of the incident photon. The left column represents the polarization of photons incident the nanophotonic platform while the top row illustrates the measured polarization. $|H\rangle$ and $|V\rangle$ represent horizontal- and vertical-linear polarization, while $|\sigma_\pm\rangle$ denote left- and right-circular polarizations **a.** Experimental results. **b.** Mean squared error fitting results for each intensity image to the superposition state of the photon in the Bessel and Bessel-Hermite bases.

Figure 2a presents the tomography measurement results for each polarization of the incident photon. Using a Mean Squared Error (MSE) estimator, we fit each measurement result to a superposition of Bessel or Hermite-Bessel modes (Fig. 2b), and infer that they all show excellent agreement with Eq. 1 and 3. Using the same MSE algorithm, we were able to reconstruct the density matrix for each of the 4 incident photons (Fig. 3). We fit each spatial distribution obtained from the tomography to the spatial modes, determining the maximum likelihood coefficients.

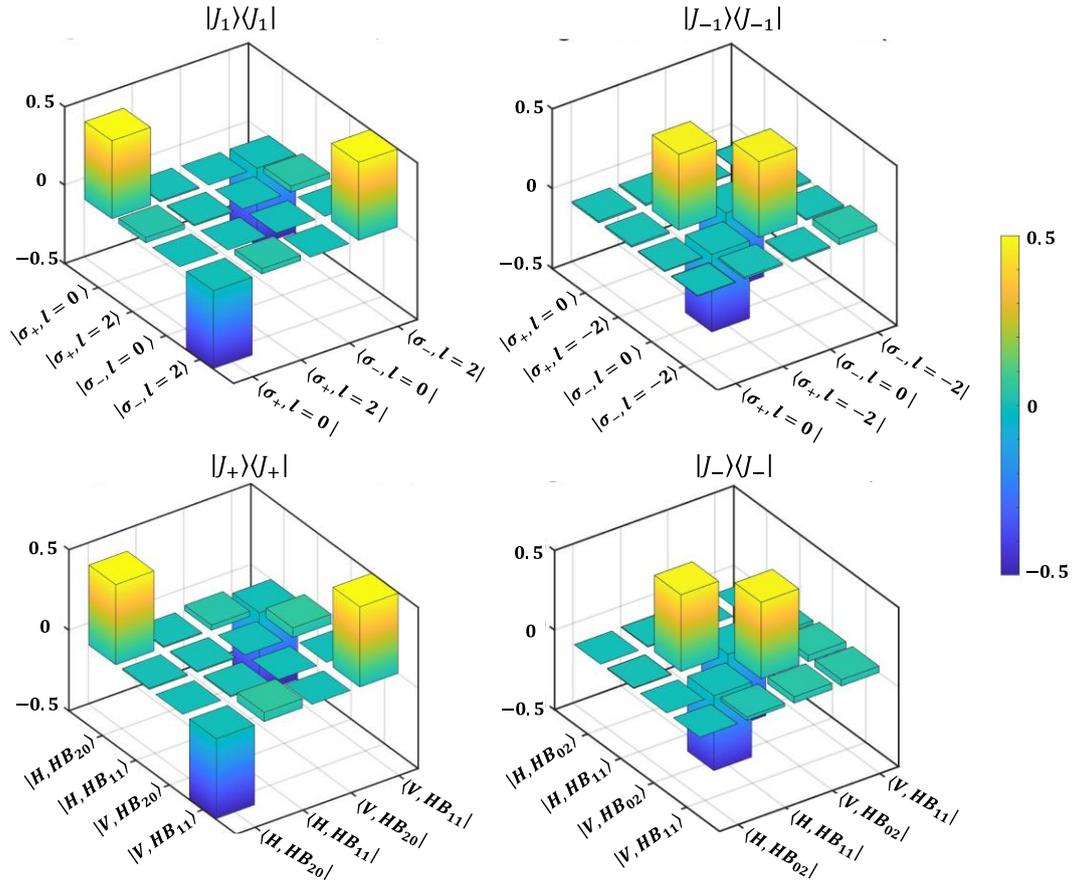

**Fig. 3. Density matrices of the four Bell states.** Experimentally measured density matrices recovered for each TAM state by the QST. The experimental results coincide with theoretical results with higher than $97 \pm 2.2\%$ fidelity. The results shown here are the real parts only because the imaginary part is identically zero both theoretically and experimentally.

To gain deeper insight into the correlations between mode and polarization (and, in particular, SAM and OAM), we calculate the Wigner distribution of the photonic mode. In fact, this distribution represents the correlation between the in-plane vector field components. Wigner distribution encodes quantum expectation values in phase-space (continuous variables), it provides a concise representation of the quantum-mechanical information in the system (see supplementary for the derivation). For each input polarization, the two entangled output modes, denoted as $\alpha$ and $\beta$, have a real and imaginary parts, making it a 4D Wigner distribution corresponding to four phase-space coordinates. We plot projections of the Wigner distribution onto six pairs of space-phase coordinates, with the two other coordinates at the origin (Fig. 4).

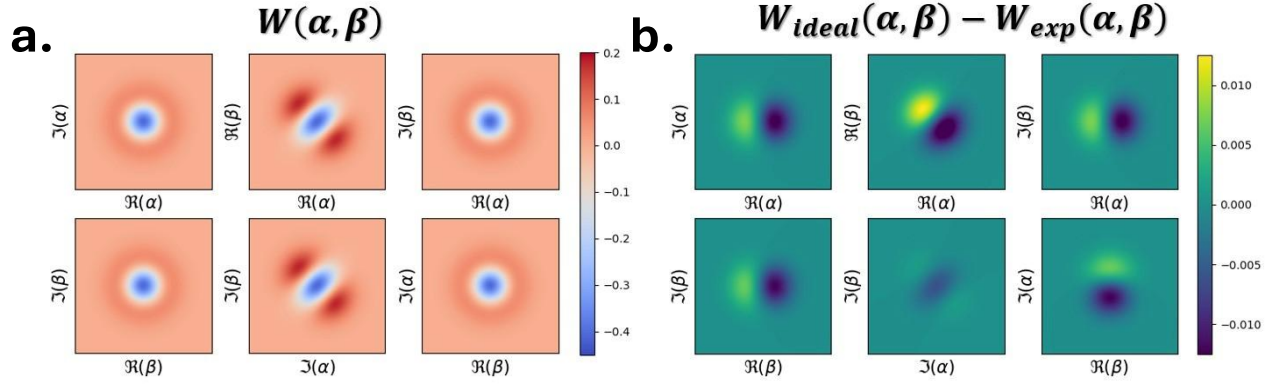

**Fig. 4. Wigner function of the photon derived from the nanophotonic mode.** The figure displays six slices of the four-dimensional Wigner function, each representing a projection onto two components of the mode. **a.** Example of the Wigner function for the state $|J_1\rangle$. **b.** The difference between the ideal Wigner function and the computed one. The various images reveal an asymmetry on the order of $10^{-2}$, which could be attributed to the partial presence of the $|J_{-1}\rangle$ state in the system's basis.

The two projections on the left-hand side of Fig. 4.a represent the phase-space of a single mode - $\Re(\alpha), \Im(\alpha)$ and $\Re(\beta), \Im(\beta)$ – each with a single excitation. The panels display negative values that attest to the non-classical nature of the state and a single ring – as expected of a Wigner distribution for the Fock state $|1\rangle$. The two projections on the right-hand side of Fig. 4.a do not exhibit correlations between the $\Re(\alpha), \Im(\beta)$ and $\Re(\beta), \Im(\alpha)$ phase-space cross-mode coordinate pairs, since they can be decomposed into a sum of products of the single-mode Wigner distributions corresponding to the Fock states $|0\rangle$ and $|1\rangle$, see supplementary. However, entanglement between the two modes is clearly evidenced in the two projections in the middle column, which exhibit nonclassical correlations between the pairs of phase-space coordinates $\Re(\alpha), \Re(\beta)$ and $\Im(\alpha), \Im(\beta)$ belonging to different modes. This can be surmised from the fact that the distribution is not symmetric about its two axes and thus cannot be the outer product of two symmetric 1D quasiprobability functions along phase-space coordinates corresponding to Fock number states 0 and 1 in the two modes, as explained in the supplementary. Our measured results agree with the theoretical Wigner distribution to an excellent degree, as seen from the differences between the theoretical and measured 2D projections of the Wigner distributions, shown in Fig. 4.b.

**Discussion and Outlook**

In conclusion, we generated entangled single-photon states using a nanophotonic platform and systematically explored the evolution of the quantum information they carry, which is associated with angular momentum, as it couples into and out of the photonic platform. These entangled states were encoded in the polarization (circular or linear) and spatial mode (Bessel or Hermite-Bessel) degrees of freedom of photons, utilizing the vector nature of TAM modes created in the near-field and the unique way in which they couple back into free-space.

Our findings could lead to a new source for non-separable quantum states, such as quantum astigmatic states (achieved through judiciously shaped couplers)[55], fractional states (using spiral couplers with fractional helicity)[56] and quantum skyrmion modes (typically with a TAM of 0 with a hexagonal coupler)[57], all of which exhibit a vector wavefunction in the near-field with embedded entanglement between their DoFs in the far-field. Additionally, the nanophotonic platform could serve as a quantum multiplexer or demultiplexer for quantum communication, enabling the entanglement of multiple photon DoFs. Our system supports quantum information encoding in the large Hilbert space of SAM and OAM qudits using simple free space OAM-selective masks and polarizers to isolate the desired states.

Our platform is well-suited for integration with a wide range of on-chip photonic circuits and could be extended to achieve higher-order entanglement by coupling multiple photons to the nanophotonic system, advancing on-chip quantum technologies. Furthermore, it offers the potential for generating novel quantum states of light with SAM and OAM by leveraging nonlinear interactions between a strong free-space pump and quantum surface-confined states, as in [58,59].

**Acknowledgements**

This research was supported by the Israeli innovation authority trough the MAGNET program, Grant number 73756, and was supported by the Israel Science Foundation (ISF), Grant number 3620/24. We acknowledge the Russell Berrie Nanotechnology Institute, Micro-Nano Fabrication Unit (MNFU) and Hellen Diller Quantum Center for their support of this research. A.K. acknowledges support by the program for graduate students in the fields of natural sciences, engineering, and medical professions by Israel Ministry of Innovation, Science, and Technology and the support from Hellen Diller Quantum Center at the Technion. S.T. acknowledges support from the Adams fellowship program of the Israel Academy of Science and Humanities, the Rothschild fellowship of the Yad Hanadiv foundation, the VATAT-Quantum fellowship of the Israel Council for Higher Education, the Helen Diller Quantum Center postdoctoral fellowship the Viterbi fellowship of the Technion – Israel Institute of Technology.

**Contributions**

G.B., M.O, S.T. and M.S. conceived the project. A.K., L.F., L.P and K.C. designed and fabricated the samples. A.K., S.T. and L.F. build the experimental platform. A.K., L.F. and S.T. conducted the measurements. A.K., A.S., A.S., G.S. and Y.I. performed simulations and analyzed the experimental results. A.K., A.S., A.S., L.N.L., S.T, and M.O. conducted the theoretical calculations. G.B., M.O. and M.S. supervised the project. All authors participated in writing the manuscript.

**Data availability**

The data supporting the findings of this study are available from the corresponding authors upon reasonable request.

**Code availability**

The codes used to process the data are available from the corresponding authors upon reasonable request.

**Ethics declarations**

Competing interests

The authors declare no competing interests.